\documentclass[twocolumn,prl,showpacs,superscriptaddress,preprintnumbers,amsmath,amssymb]{revtex4}

\usepackage{graphicx}% Include figure files
\usepackage{dcolumn}% Align table columns on decimal point
\usepackage{bm}% bold math
\usepackage{color}
\usepackage{xfrac}
\usepackage{tikz}
\usetikzlibrary{shapes.geometric,arrows,positioning,calc,decorations.markings}

\begin{document}
\newcommand{\siox}{SiO$_2$}
\newcommand{\silicate}{Si$_2$O$_3$}
\newcommand{\sicoxint}{\mbox{SiC/SiO$_2$}}
\newcommand{\rootthree}{($\sqrt{3}$$\times$$\sqrt{3}$)R30$^{\circ}$}
\newcommand{\rootthreehl}{($\sqrt{\mathbf{3}}$$\times$$\sqrt{\mathbf{3}}$)R30$^{\circ}$}
\newcommand{\sixroot}{(6$\sqrt{3}$$\times$6$\sqrt{3}$)R30$^{\circ}$}
\newcommand{\sixroothl}{(6$\sqrt{\mathbf{3}}$$\times$6$\sqrt{\mathbf{3}}$)R30$^{\circ}$}
\newcommand{\three}{\mbox{(3${\times}$3)}}
\newcommand{\four}{\mbox{(4${\times}$4)}}
\newcommand{\five}{\mbox{(5${\times}$5)}}
\newcommand{\six}{\mbox{(6${\times}$6)}}
\newcommand{\two}{\mbox{(2${\times}$2)}}
\newcommand{\twobyc}{\mbox{(2$\times$2)$_\mathrm{C}$}}
\newcommand{\twobysi}{\mbox{(2$\times$2)$_\mathrm{Si}$}}
\newcommand{\seven}{($\sqrt{7}$$\times$$\sqrt{7}$)R19.1$^{\circ}$}
\newcommand{\fourseven}{(4$\sqrt{7}$$\times$4$\sqrt{7}$)R19.1$^{\circ}$}
\newcommand{\one}{\mbox{(1${\times}$1)}}
\newcommand{\sicbar}{SiC(000$\bar{1}$)}
\newcommand{\sicbarhl}{SiC(000$\bar{\mathbf{1}}$)}
\newcommand{\bardir}{(000$\bar{1}$)}
\newcommand{\grad}{\mbox{$^{\circ}$}}
\newcommand{\cgrad}{\,$^{\circ}$C}
\newcommand{\projecta}{(11$\bar{2}$0)}
\newcommand{\projectb}{(10$\bar{1}$0)}
\newcommand{\projectc}{(01$\bar{1}$0)}
\newcommand{\third}{$\frac{1}{3}$}
\newcommand{\thirdspot}{($\frac{1}{3}$,$\frac{1}{3}$)}
\newcommand{\oversix}{$\frac{1}{6}$}
\newcommand{\gkdir}{$\overline{\Gamma \mbox{K}}$-direction}
\newcommand{\pb}{$\pi$ band}
\newcommand{\pstar}{$\pi$* band}
\newcommand{\pbs}{$\pi$ bands}
\newcommand{\kv}{\mathbf{k}}
\newcommand{\kpoint}{$\overline{\textrm{K}}$ point}
\newcommand{\mpoint}{$\overline{\textrm{M}}$ point}
\newcommand{\kpp}{$\overline{\textrm{K'}}$}
\newcommand{\kp}{$\overline{\textrm{K}}$}
\newcommand{\mbar}{$\overline{\textrm{M}}$}
\newcommand{\gpoint}{$\overline{\textrm{\Gamma}}$ point}
\newcommand{\gk}{$\overline{\Gamma \mbox{K}}$}
\newcommand{\kpar}{$\mathbf{k}_{\parallel}$}
\newcommand{\efermi}{E$_\mathrm{F}$}
\newcommand{\edirac}{E$_\mathrm{D}$}
\newcommand{\angs}{$\mathrm{\AA}$}
\newcommand{\kk}  {$\mathbf{k}_{\overline{\textrm{K}}}$}
\newcommand{\kvec}{\underline{k}}
\newcommand{\sbs}[2]{\rlap{\textsuperscript{{#1}}}\textsubscript{#2}}
\newcommand{\round}[1]{\left({#1}\right)}
\newcommand{\moire}{moir{\'e}}
\newcommand{\gm}{$\overline{\Gamma}$}

\newcommand{\red}{\textcolor[rgb]{1,0,0}}

\newcommand{\todo}[1]{\textsl{\textcolor{red}{#1}}}

\newcommand{\kay}{\mathbf{k}}
\newcommand{\cue}{\mathbf{q}}
\newcommand{\up}{\uparrow}
\newcommand{\down}{\downarrow}
\newcommand{\str}{^{*}}
\newcommand{\pathI}{\int D[c\str,c]}

\title{Introducing strong correlation effects into graphene by gadolinium intercalation}

\author{S. Link}
\affiliation{Max-Planck-Institut f\"{u}r Festk\"{o}rperforschung, Heisenbergstr. 1, D-70569 Stuttgart, Germany}

\author{S. Forti}
\altaffiliation[present address:~]{Center for Nanotechnology Innovation CNI@NEST, Piazza San Silvestro 12, 56127 Pisa, Italy}
\affiliation{Max-Planck-Institut f\"{u}r Festk\"{o}rperforschung, Heisenbergstr. 1, D-70569 Stuttgart, Germany}

\author{A. St\"{o}hr}
\affiliation{Max-Planck-Institut f\"{u}r Festk\"{o}rperforschung, Heisenbergstr. 1, D-70569 Stuttgart, Germany}

\author{K. K\"{u}ster}
\affiliation{Max-Planck-Institut f\"{u}r Festk\"{o}rperforschung,
Heisenbergstr. 1, D-70569 Stuttgart, Germany}

\author{M. R\"{o}sner}
\affiliation{Institute for Theoretical Physics, Bremen Center for
Computational Materials Science, University of Bremen Otto-Hahn-Allee 1,
28359 Bremen, Germany} \affiliation{Department of Physics and Astronomy,
University of Southern California, Los Angeles, CA 90089-0484, USA}
\affiliation{Radboud University, Institute for Molecules \& Materials,
Heijendaalseweg 135, 6525 AJ Nijmegen, Netherlands}

\author{D. Hirschmeier}
\affiliation{Universit\"{a}t Hamburg, Institut f\"{u}r Theoretische Physik, D-20355 Hamburg, Germany}

\author{C. Chen}
\affiliation{Synchrotron SOLEIL and Universit\'e Paris Saclay, Orme des Merisiers, Saint-Aubin-BP 48, 91192 Gif sur Yvette, France}

\author{J. Avila}
\affiliation{Synchrotron SOLEIL and Universit\'e Paris Saclay, Orme des Merisiers, Saint-Aubin-BP 48, 91192 Gif sur Yvette, France}

\author{M.C. Asensio}
\affiliation{Madrid Institute of Materials Science (ICMM), Spanish Scientific
Research Council (CSIC), Cantoblanco, E-28049 Madrid - SPAIN}

\author{A.A. Zakharov}
\affiliation{MAX IV Laboratory, Lund University, Fotongatan 2, 22484 Lund,
Sweden}

\author{T.O. Wehling}
\affiliation{Institute for Theoretical Physics, Bremen Center for
Computational Materials Science, University of Bremen Otto-Hahn-Allee 1,
28359 Bremen, Germany}

\author{A.I. Lichtenstein}
\affiliation{Universit\"{a}t Hamburg, Institut f\"{u}r Theoretische Physik, D-20355 Hamburg, Germany}

\author{M.I. Katsnelson}
\affiliation{Radboud University, Institute for Molecules \& Materials, Heijendaalseweg 135, 6525 AJ Nijmegen, Netherlands}

\author{U. Starke}%
 \email[To whom correspondence should be addressed:\\ Email: ]{u.starke@fkf.mpg.de}
 \homepage[\\Electronic adress: ]{http://www.fkf.mpg.de/ga}
\affiliation{Max-Planck-Institut f\"{u}r Festk\"{o}rperforschung, Heisenbergstr. 1, D-70569 Stuttgart, Germany}

\date{\today}

%%%%%%%%%%%%%%%%% END OF PREAMBLE %%%%%%%%%%%%%%%%

\begin{abstract}
Exotic ordered ground states driven by electronic correlations are expected
to be induced in monolayer graphene when doped to the Van Hove singularity.
Such doping levels are reached by intercalating Gd in graphene on SiC(0001),
resulting in a strong homogeneity and stability. The electronic spectrum now
exhibits severe renormalizations. Flat bands develop which is driven by
electronic correlations according to our theoretical studies. Due to strong
electron-phonon coupling in this regime, polaron replica bands develop.
Thus, interesting ordered ground states should be made accessible.
\end{abstract}

\maketitle

Monolayer graphene can be viewed as prototype of the materials' class of
Dirac materials~\cite{Novoselov2005,Zhang2005,Castro2009,Berger2006}. At low
carrier concentrations, the low energy excitations can be described as
massless Dirac Fermions, as the electronic spectrum is nominally linear in a
single particle picture, resulting in a conical shape at the high symmetry
points {\kp} and {\kpp} of graphene's Brillouin zone (BZ) (see
Fig.~\ref{fig1} (a)). However, when including many body interactions,
especially electronic correlations, the bands can be reshaped drastically,
due to the strong onsite Coulomb repulsion in graphitic
systems~\cite{Wehling2011}. Furthermore, long range electronic correlations
are substantial in this system due to its two-dimensionality and Dirac-like
spectrum~\cite{Sarma2007,Bostwick2007}. At low doping levels, this results in
logarithmic corrections in the electronic
spectrum~\cite{Siegel2011,Elias2011}. At finite n-type doping levels, the
total band width of the $\pi$ band structure gets significantly
lowered~\cite{Ulstrup2016}. In the extreme case of high doping, a Van Hove
singularity (VHs) in the density of states is reached, whereas flat bands
develop at the Fermi level ($E_F$), connecting {\kp} and
{\kpp}~\cite{McChesney2010}. Similar effects were observed in strongly
correlated materials like cuprates~\cite{Gofron1994} and
ruthenates~\cite{Lu1996}, referred to as extended Van Hove singularities
(eVHs) \cite{Khodel1990,Nozieres1992,Irkhin2002,Yudin2014}. In the case of
cuprates, this then called Van Hove (VH) scenario appears to be partially
connected to the rich and unusual physical phenomena and to the different
phases found in those materials~\cite{Bok2012}. Despite the fundamental
differences to these systems, one might expect to find similarly rich physics
in graphene in this regime. Generically, the emergence of an eVHs is a
hallmark of a non-Fermi liquid state of matter called Fermi condensate
\cite{Khodel1990,Nozieres1992,Irkhin2002,Yudin2014} where the electronic
dispersion becomes flat at the chemical potential in a finite {momentum
interval $k_{1} < |k| < k_{2}$, i.e. $e_{k}$ = $\mu$ for T $\rightarrow$ 0.
In this light, it is especially interesting to draw the connection to
predictions of exotic ordered ground states in graphene at VH-filling, like
chiral superconductivity driven by repulsive electron-electron
interaction~\cite{Nandkishore2012,Kiesel2012}, where such strong band
renormalizations are not taken into account.

In order to experimentally reach the VHs in graphene, extreme carrier
induction is required which can only be achieved by chemical doping. For this
purpose a combination of adsorption and intercalation of Ca and K was applied
before~\cite{McChesney2010}, where the surface exposure and reactivity of
dopants cause severe limitations in thermal and chemical stability.
Intercalation alone, however, i.e., the full insertion of the dopant between
graphene and its substrate assures a physical (and chemical) protection by
the covering graphene layer~\cite{Riedl2009,Gierz2010}. Thus, detailed
in-situ band structure analysis~\cite{Forti2DMat}, characterization with
ex-situ techniques~\cite{Gierz2010} or further processing into a device are
then possible~\cite{Baringhaus2015}. Indeed, transfer and contacting under Ar
atmosphere allows the application of cryogenic transport experiments, as
currently underway.

Accordingly, here we intercalate Gd atoms beneath the so called zero-layer
graphene (ZLG) on SiC(0001)~\cite{supp}.
Intercalation was performed by depositing Gd from an e-beam evaporator
combined with heating up to temperatures of 1200 {\cgrad}. The
Gd-intercalated graphene is thermally stable and sufficiently doped to
provide VH-filling. The intercalation geometry (see Fig.~\ref{fig1} (b))
provides chemical stability, such that no degeneration in the samples could
be observed within weeks, when kept under ultra-high vacuum conditions
($p<~1\times 10^{-8}$~mbar) in contrast to ref.~\onlinecite{McChesney2010}
where doping from both sides makes the system severly environment sensitive.
Low energy electron microscopy (LEEM) experiments prove the homogeneity on a
mesoscopic scale~\cite{supp}. LEEM as well as X-ray photoelectron
spectroscopy (XPS) were carried out at beamline I311 of MAX-lab synchrotron,
Lund (Sweden). The stability supports high quality band structure
measurements using angle-resolved photoelectron spectroscopy (ARPES) which
was conducted at beamline ANTARES of SOLEIL synchrotron (France), at Bessy II
beamline UE112 of Helmholtz Zentrum Berlin and at beamline I4 of MAX-Lab.

The experimental results are simulated by band modeling within density
functional theory (DFT) for Gd adsorbed on graphene~\cite{supp}, which can reproduce the observed doping levels, yet cannot account for the strong
renormalizations found in the experiment. The DFT calculations were performed
within the generalized gradient approximation (GGA) \cite{Perdew1996} using
the Vienna Ab Initio Simulation Package (VASP) \cite{Kresse1994} and the
projector augmented wave (PAW) \cite{Bloechel1994, Kresse1999} basis sets. To
overcome the limitations of DFT based mean field techniques, we utilize the
Hubbard model, which is solved in fluctuating exchange approximation (FLEX)
\cite{Bickers1989,supp}. Introducing the effects of spin fluctuations, we
could reproduce the experimental results. The temperature was chosen to be
$T=33\,$meV ($386\,$K) in all calculations.

Fig.\ \ref{fig1}(c) displays the result of ARPES measurements of Gd
intercalated ZLG on SiC(0001) spanning a cut from {\kp} over {\mbar} to
{\kpp}~\cite{supp,ARPESfootnote}.
\begin{figure}[t!]
	\begin{center}
		\includegraphics[width=0.47\textwidth]{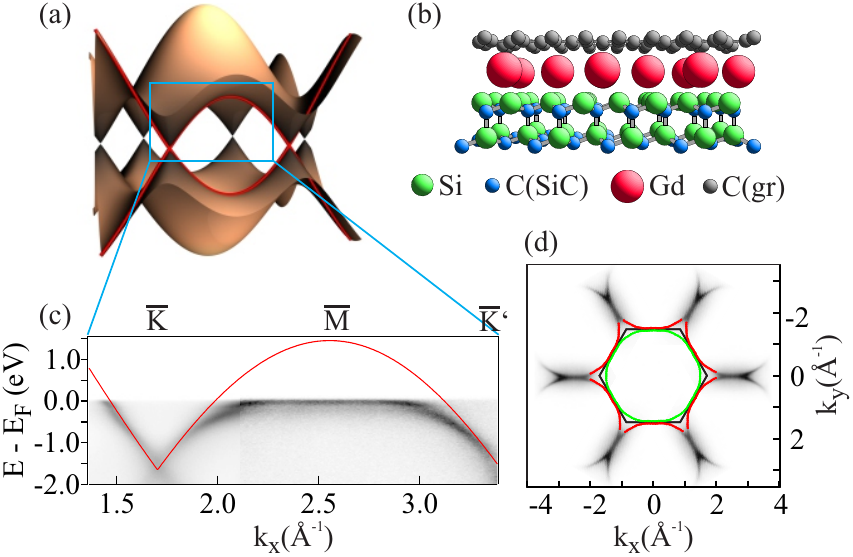}
	\end{center}
	\caption{ARPES on Gd intercalated ZLG: (a) Illustration of graphene's $\pi$ bands. (b) Side view model of
		the intercalation system. (c)
		Two concatenated ARPES measurements cutting from {\kp} over
		{\mbar} to {\kpp} together with a band modeled with NN-TB (red
		trace). The left part (below 2.1~\AA$^{-1}$) was taken with 30~eV and the right part (above 2.1~\AA$^{-1}$) with 100~eV
		photon energy. (d) Symmetrized FS taken with 90~eV photon energy
		together with its experimental fit (red and green lines). The black hexagon represents graphene's first BZ.}
	\label{fig1}
\end{figure}
Around {\kp}, the conical $\pi$ band of monolayer graphene emerges, which
proves successful intercalation~\cite{Riedl2009,supp}. In contrast to other
intercalants, Gd produces graphene with an extremely high n-type doping
level. The Dirac point is shifted by about 1.6~$\textup{eV}$ from $E_F$ to
higher binding energy. This is sufficient to reach the saddle point at
{\mbar}, as seen by the strong spectral weight at {\mbar} near $E_F$ -- now
by intercalation alone. Concomitantly, the Fermi surface (FS) experiences a
Lifshitz transition, cf.\ Fig.\ \ref{fig1}(d). Instead of two electron
pockets, centered around {\kp} and {\kpp}, one giant hole pocket emerges
around {\gm}. From an experimental fit to the FS, one can directly calculate
a filling of about 0.12 in the {\pstar} following Luttinger's
theorem~\cite{Luttinger1960}. This corresponds to about $4.5\times
10^{14}/\textup{cm}^2$ additional electrons compared to neutral
graphene~\cite{FScalculation}. In view of a single electron picture
(nearest-neighbor tight-binding, NN-TB), this would not be sufficient to
reach the VH-filling. There the transition should be at 1/4 filling or
$9.5\times 10^{14}/\textup{cm}^2$ electron density. Further differences can
be seen in the dispersion in Fig.\ \ref{fig1}(c). While the bands along the
{\gk}-direction can be fitted well with NN-TB bands with a reasonable hopping
parameter of 3.1~eV, the bands along the {\kp}{\mbar}-direction deviate
strongly. Near {\kp} the band velocity is already drastically reduced, whilst
around {\mbar} the band is completely flat over a range of about 1~\AA$^{-1}$
with renormalizations on the order of 1.5~eV as compared to the NN-TB model
(red trace in the figure).
\begin{figure}[ht!]
	\begin{center}
		\includegraphics[width=0.47\textwidth]{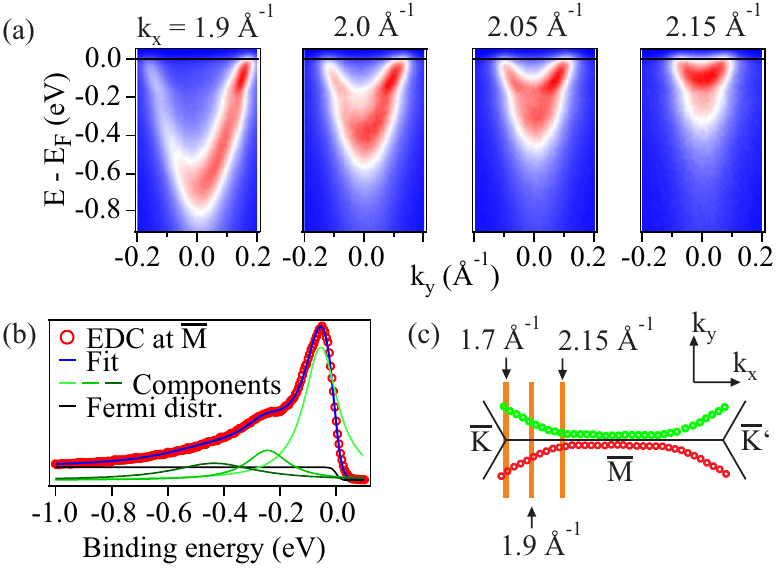}
	\end{center}
	\caption{ Electron-phonon coupling around {\mbar}: (a) Series of ARPES
cuts taken perpendicular to the {\kp}{\mbar}-direction at different k$_x$,
as indicated in panel (c), photon energy 35 eV. (b) EDC at {\mbar} together with
a fit of 3 equidistant Lorentzians multiplied with the Fermi
distribution. (c) Fitted FS around {\mbar}, illustrating the
parallel sections of the FS.}
	\label{fig2}
\end{figure}
Optical absorption experiments taken on doped graphene~\cite{Mak2014}
revealed the saddle point at {\mbar} also at significantly lower binding
energy compared to NN-TB. The binding energy reduction increased with the
doping level, which was associated with electron-electron interaction.
However, the situation in these experiments is still distinctly different
from a flat band near {\mbar}, which cannot be accounted for by any smooth
modification to NN-TB.

As apparent in Fig.\ \ref{fig1}(c), the flat band at {\mbar} is accompanied
by a second band at higher binding energy, as also observed in
ref.~\cite{McChesney2010}. This spectral feature can be associated with
electron-boson coupling, as illustrated further in Fig.\ \ref{fig2}(a) by
spectral cuts taken perpendicular to the cut in Fig.\ \ref{fig1}(c),
revealing the spectral evolution from {\kp} towards {\mbar} of the typical
kink structure near $E_F$, induced by phonon coupling. At about
$k_x$=2~\AA$^{-1}$, the kink structure evolves smoothly into a structure,
where a sharp coherent branch near $E_F$ is replicated by a broader
incoherent part at higher binding energy (second band). Taking an energy
distribution curve (EDC) from {\mbar} (see Fig.\ \ref{fig2}(b)), the
distribution can be fitted by three Lorentzians with a mutual distance of
190~meV. This energy corresponds to graphene's in-plane optical phonon modes
at {\gm}. The observed spectral feature is thus likely associated with the
formation of polarons as new electron-phonon composite
quasiparticles~\cite{Mishchenko2000}. From the spectral fraction of the
coherent part near $E_F$ of 0.6, one can estimate the coupling constant
$\alpha$ to this particular phonon mode to be on the order of 1 by comparing
to diagrammatic Monte-Carlo simulations~\cite{Mishchenko2000}. This is
strongly enhanced compared to coupling constants to this phonon mode in
graphene, which is not in the eVHs regime~\cite{Fedorov2014}. A detailed,
zoomed view of the FS, cf.\ Fig.\ \ref{fig2}(c), demonstrates that the sector
at {\mbar} is very straight and thereby nested to its counterpart in the
repeated BZ. Generally, this condition strongly enhances resonant coupling,
as the phase space for phonon scattering within these sections of the FS is
enhanced. As the two FS sections are very close, the corresponding phonons
have about zero momentum, i.e. they are located around {\gm} in the BZ. The
previously extracted energy of the phonon mode is in agreement with this
picture. These phonon-induced spectral features may have a strong influence
on or even enhance the observed flat band section. However, with reasonable
coupling constants, this effect alone cannot induce the flat band.

In order to provide information about interface structure and chemistry,
low-energy electron diffraction (LEED) and XPS experiments were conducted.
Fig.\ \ref{fig3}(a) displays two LEED patterns taken with different electron
energies.
\begin{figure}[t!]
	\begin{center}
		\includegraphics[width=0.47\textwidth]{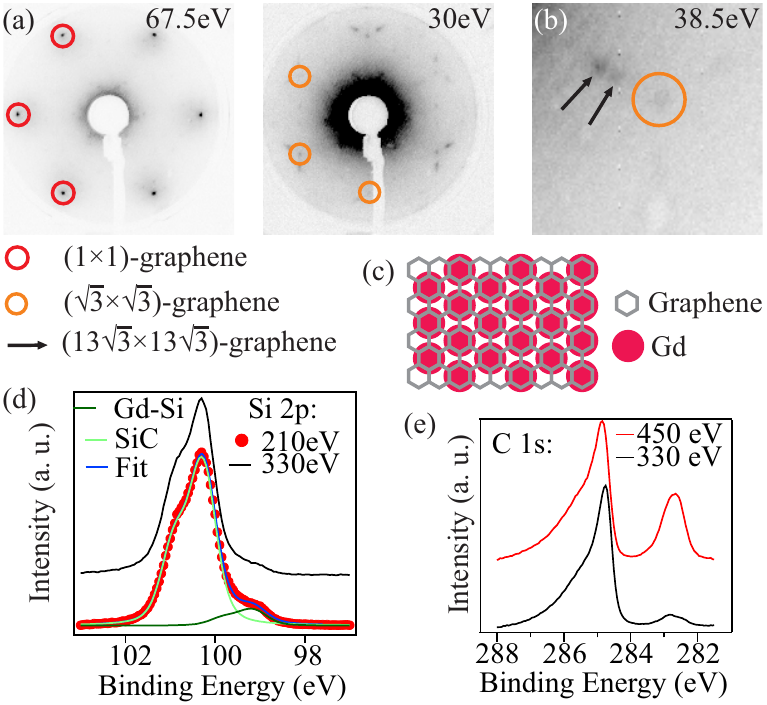}
	\end{center}
	\caption{ LEED and XPS on Gd intercalated ZLG: (a) LEED patterns taken with
		two different electron energies. (b) Diffraction pattern closeup around a graphene
		{\thirdspot}-spot. The different spots in (a,b,c) are ascribed in the legend.
        (c) Sketch of a simplified Gd-($\sqrt{3}$$\times$$\sqrt{3}$)$\textup{R}30^\circ$-graphene structure.
        (d) Si~2p spectra of Gd intercalated ZLG taken with two different photon energies together with a
        fit of the two involved chemical species. (e) C~1s spectra of Gd intercalated ZLG
        taken with two different photon energies. }
	\label{fig3}
\end{figure}
As expected for intercalated ZLG, none of the reconstruction spots associated
with the pristine ZLG structure are visible, since the covalent bonding to
the substrate in the pristine ZLG system is lifted. Instead, the first order
(1$\times$1)-spots of graphene emerge sharp and bright (67.5~eV pattern).
Besides, a new periodicity can be recognized. As apparent in the 30~eV
pattern by the marked {\thirdspot}-spots, this periodicity is based on a
{\rootthree} superstructure with respect to graphene. Since the
{\rootthree}-graphene is not part of the total periodicity of the initial ZLG
system ((13$\times$13)-graphene), we conclude that the intercalated Gd
primarily arranges with respect to the graphene. Note that the amount of
deposited and intercalated Gd, i.e. 0.75-0.8 monolayer of Gd~\cite{ML-note},
corresponds to a simple {\rootthree}-graphene structure, as sketched in Fig.\
\ref{fig3}(c). Yet, it is apparent especially from the elongated spots around
the {\thirdspot}-spots, that this structure itself is reconstructed. In fact,
the elongated features around the {\thirdspot}-spots are two very close
spots, as seen in the closeup in Fig.\ \ref{fig3}(b). Also, an apparent
threefold symmetry in the varying intensity of these {\thirdspot}-spots can
be recognized which cannot be explained by the Gd-graphene slab alone.

It is intuitive to explain these modifications by the influence of the
substrate in the form of a bonding of the Gd to the topmost Si of the
SiC(0001) surface. This is corroborated by the fact that in the Si~2p spectra
in XPS (see Fig.\ \ref{fig3}(d)), two different chemical species of Si can be
distinguished. Varying the photon energy and	thereby surface sensitivity
shows that the doublet around 100.5~eV has to be associated with Si in the
bulk SiC and that at about 99~eV with Si at the SiC surface, since the latter
is more pronounced in the more surface sensitive probe (210~eV photon
energy). The strong energy shift gives evidence for a charge transfer and
thus for a bonding to Gd. By comparing the intensity ratio of the two
chemical species (5:95) to the case of H-intercalated ZLG
(1:2)~\cite{Riedl2009}, where every Si on the surface is passivated by one H
atom, it can be inferred that only about 3/20 of the Si atoms are passivated
in the Gd case. Taking into account the amount of intercalated Gd and the
assumption that only one Gd atom is bound to a Si atom respectively, one can
estimate, that about 1/3 of the Gd atoms bind to Si. Connecting these results
further to the observations made with LEED, the partial bonding explains the
apparent threefold symmetry, which originates from the stacking order of the
6H-SiC substrate. Also, as the total periodicity of the graphene-SiC system
is a (13$\times$13)-graphene, one can expect for the total system a
($\sqrt{3}$$\times$$\sqrt{3}$)$\textup{R}30^\circ$-superstructure on the
(13$\times$13)-graphene, i.e. a
($13\sqrt{3}$$\times$$13\sqrt{3}$)$\textup{R}30^\circ$-graphene. Indeed, this
model captures all the observed diffraction spots.

Concerning the chemistry within the graphene layer, Fig.\ \ref{fig3}(e)
displays C~1s spectra taken with two different photon energies. Two prominent
main features can be determined, that must be associated with bulk SiC around
282.5~eV and the graphene between 284.5~eV and 288~eV since in the more bulk
sensitive probe (450~eV photon energy), the 282.5~eV binding energy component
is clearly more pronounced. In contrast to the low doped case, the C~1s
component of graphene does not follow a simple asymmetric distribution that can
be simulated by a Doniach Sunjic function, which is often suitable for metals.
This is apparent in the clear hump at about 285.8~eV binding energy. Simulations
in ref.~\cite{Sernelius2015} and also experimental work in ref.~\cite{Schroder2017}
for highly doped graphene corroborate our findings. The characteristic shape is meant
to be induced by energy loss primarily due to plasmon creation during the photoemission process.

%
% Theory
%
%
\begin{figure}[t!]
	\begin{center}
		\includegraphics[width=0.47\textwidth]{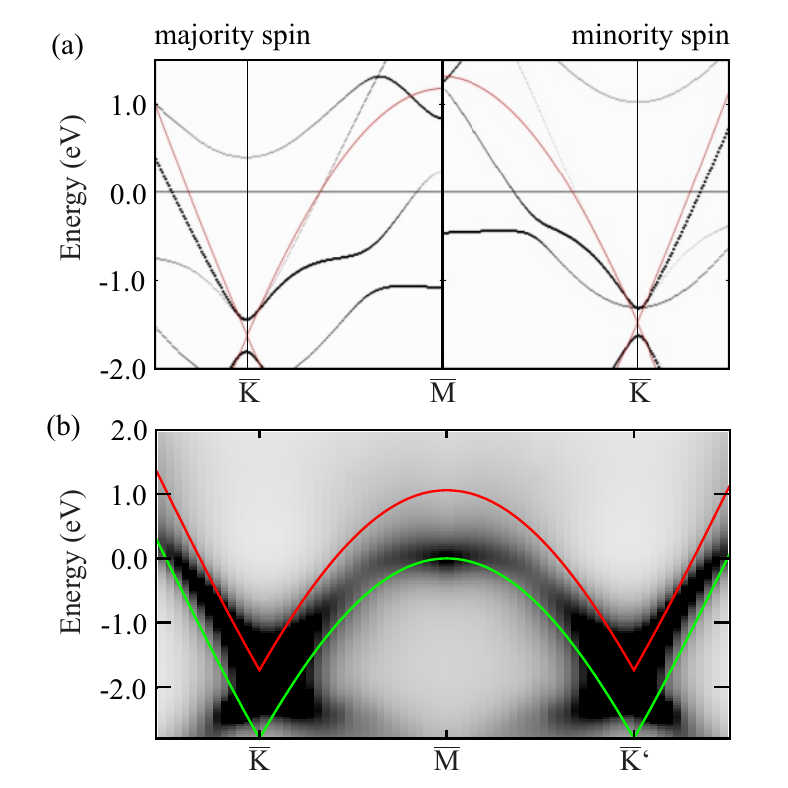}
	\end{center}
	\caption{(a) DFT band structure of Gd on
graphene in the ($\sqrt{3}\times\sqrt{3}$)R30$^\circ$ structure unfolded to
the BZ of the primitive graphene unit cell (gray lines) and
NN-TB band structure (red). The line thickness of
the unfolded bands serves as an approximate measure of an intensity expected
in an ARPES experiment. Precisely, the line thickness at some wave vector $k$
from the BZ of the orginal graphene lattice indicates the overlap
of the corresponding state in the ($\sqrt{3}\times\sqrt{3}$)R30$^\circ$ supercell
with a plane wave from the primitive graphene unit cell at wave vector $k$.
b) Spectral function as obtained from fluctuating exchange approximation
theory. The essential features of the ARPES data (eVHs and the energetic
shift of the renormalized spectrum at the Dirac point) are well reproduced.
Comparison to the NN-TB model at VH-filling (green curve) as well as a
NN-TB model fit to the shift at the {\kp} point (red) implies that the
renormalization is correlation induced.}
	\label{fig4}
\end{figure}
In order to reveal the mechanism responsible for the flat band formation,
we performed DFT simulations of Gd adsorbed on graphene in the
{\rootthree} structure sketched in Fig. \ref{fig3}(c) and calculated the
upfolded band structure. The results shown in Fig. \ref{fig4}(a) reveal an
essentially intact graphene Dirac point, located approximately $1.5\,$eV below $E_F$,
which is in line with our ARPES experiments. Hybridization with Gd bands pushes
graphene's $\pi$ bands at {\mbar} more than $0.5\,$eV below $E_F$ and induces a
spin-splitting of about $0.5\,$eV. While a downward shift of the bands at {\mbar}
as compared to NN-TB is in line with our experimental observation, the magnitude
of the shift in DFT largely exceeds the experimental one. In particular,
there is no pinning of any flat band to $E_F$ in the {\kp}-{\mbar} direction
in the DFT results. Indeed, such pinning is a typical hallmark of dynamic
electron correlation effects \cite{Yudin2014}, which are beyond DFT based mean field theories.
To investigate whether these electronic correlations in form of spin-fluctuations
can be responsible for the flattening of the bands along {\kp}-{\mbar},
we considered the following Hubbard model
\begin{equation}\label{eq:Hubbard}
H=-t\sum_{\langle i,j\rangle,\sigma} c^{\dagger}_{i\sigma}c_{j\sigma}-\mu\sum_{i,\sigma}c^{\dagger}_{i\sigma}c_{i\sigma}+U\sum_{i}n_{i,\uparrow}n_{i,\downarrow},
\end{equation}
where electrons on site $i$ with spin $\sigma$ are created (annihilated) by
$c^{\dagger}_{i\sigma}$ ($c_{i\sigma}$) and the density operator is given by
$n_{i\sigma}=c^{\dagger}_{i\sigma}c_{i\sigma}$. The nearest-neighbour hopping
amplitude in our model calculations is $t=2.8\,$eV, while the chemical
potential is set to $\mu=2.4\,$eV and the Hubbard repulsion is chosen to be
$U=8.4\,$eV. We solve this model using the FLEX approximation
\cite{Bickers1989} which introduces the effects of spin fluctuations. The
resulting FLEX spectral function is shown in Fig. \ref{fig4}(b). The
comparison with a NN-TB model at VH-filling (green curve) clearly shows that
the VHs at $E_F$ is extended to a flat band that almost stretches half way to
{\kp}. Furthermore, the spectral weight at {\kp} is shifted upwards such that
the energy difference of the spectrum between {\kp} and {\mbar} is
significantly reduced to $t/2$ (while it is $t$ in the non-interacting NN-TB
model). This redistribution of spectral weight can not be described by
shifted one-particle NN-TB models (compare red and green curves) which
suggests that strong electronic correlations in form of spin fluctuations
present indeed a possible origin of the eVHs. Like in the ARPES spectral
function, there are no severe renormalizations along the {\gm}--{\kp} path,
resulting in a good agreement with the experiment. In contrast to the
experimental spectral function we do not see any indications of a second band
close to (below) $E_F$. The experimentally observed accompanying band is thus
indeed arising from electron-phonon interactions and not from interactions
with magnetic fluctuations like paramagnons.

In summary, we have shown that by intercalating Gd underneath zero-layer
graphene on SiC(0001) the carbon layer can be decoupled from its substrate.
Thereby a strongly doped monolayer graphene system is produced, where an
eVHs emerges in the spectral function indicating the realization of a Fermi
condensate in graphene. By comparing the experimental spectral function to
theoretical results from super-cell DFT as well as many-body FLEX
calculations, we found that strong electron-electron correlation effects in form
of spin-fluctuations can explain the occurrence of the eVHs, while coupling to
phonons seems responsible for the replicas of the eVHs bands at higher binding energies.

We have shown that the right choice of intercalant can result in very high thermal
and chemical stability with superb crystallinity. Together with the homogeneity on a
large scale, this gives the opportunity to study graphene's many-body properties in
the eVHs regime in great detail.
Our findings enable the stable and precisely controlled functionalization of graphene,
driving the confluence of important technological applications with the quest for
microscopic mechanisms that control the properties of the material.

\begin{acknowledgments}
We are indebted to MAX-Lab (Lund, Sweden), SOLEIL (Gif-sur-Yvette, France)
and BESSY (Berlin, Germany) for allocating synchrotron beamtime and the
facility staff for their advice and support. We would like to thank M. Konuma
for his help in XPS measurements. In addition, we would like to acknowledge
T. Denig for fruitful discussions in the initial stages of the project. This
work was supported by the German Research Foundation (DFG) in the framework
of the Priority Program 1459, Graphene. Support by the DFG through Sta315/9-1
is acknowledged in addition. M.R. would like to thank the Alexander von
Humboldt Foundation for support. A.A.Z. would like to acknowledge support
from Siftelsen f\"{o}r Strategisk Forskning (project RMA15-0024). M.I.K.
acknowledges financial support by NWO via the Spinoza Prize. Computational
resources were provided by the HLRN-Cluster under Project No. hhp00042.
\end{acknowledgments}

%\newpage

\newpage
\clearpage


\begin{thebibliography}{10}
\bibitem{ARPESfootnote} Two different photon energies were
    used (30 eV and
    100 eV) to access both the inner {\pstar} and the band course up to the
    outer {\kpp} (see \cite{supp} for details and the original data sets).
\bibitem{FScalculation} The electron density is derived from
    the difference
    between the area of the entire BZ and the hole pocket.
\bibitem{ML-note} 1 ML is defined as a one atom thick layer of the (0001)
    plane of Gd.
\end{thebibliography}
\end{document}

\newcommand{\siox}{SiO$_2$}
\newcommand{\silicate}{Si$_2$O$_3$}
\newcommand{\sicoxint}{\mbox{SiC/SiO$_2$}}
\newcommand{\rootthree}{($\sqrt{3}$$\times$$\sqrt{3}$)R30$^{\circ}$}
\newcommand{\rootthreehl}{($\sqrt{\mathbf{3}}$$\times$$\sqrt{\mathbf{3}}$)R30$^{\circ}$}
\newcommand{\sixroot}{(6$\sqrt{3}$$\times$6$\sqrt{3}$)R30$^{\circ}$}
\newcommand{\sixroothl}{(6$\sqrt{\mathbf{3}}$$\times$6$\sqrt{\mathbf{3}}$)R30$^{\circ}$}
\newcommand{\three}{\mbox{(3${\times}$3)}}
\newcommand{\four}{\mbox{(4${\times}$4)}}
\newcommand{\five}{\mbox{(5${\times}$5)}}
\newcommand{\six}{\mbox{(6${\times}$6)}}
\newcommand{\two}{\mbox{(2${\times}$2)}}
\newcommand{\twobyc}{\mbox{(2$\times$2)$_\mathrm{C}$}}
\newcommand{\twobysi}{\mbox{(2$\times$2)$_\mathrm{Si}$}}
\newcommand{\seven}{($\sqrt{7}$$\times$$\sqrt{7}$)R19.1$^{\circ}$}
\newcommand{\fourseven}{(4$\sqrt{7}$$\times$4$\sqrt{7}$)R19.1$^{\circ}$}
\newcommand{\one}{\mbox{(1${\times}$1)}}
\newcommand{\sicbar}{SiC(000$\bar{1}$)}
\newcommand{\sicbarhl}{SiC(000$\bar{\mathbf{1}}$)}
\newcommand{\bardir}{(000$\bar{1}$)}
\newcommand{\grad}{\mbox{$^{\circ}$}}
\newcommand{\cgrad}{\,$^{\circ}$C}
\newcommand{\projecta}{(11$\bar{2}$0)}
\newcommand{\projectb}{(10$\bar{1}$0)}
\newcommand{\projectc}{(01$\bar{1}$0)}
\newcommand{\third}{$\frac{1}{3}$}
\newcommand{\thirdspot}{($\frac{1}{3}$,$\frac{1}{3}$)}
\newcommand{\oversix}{$\frac{1}{6}$}
\newcommand{\gkdir}{$\overline{\Gamma \mbox{K}}$ direction}
\newcommand{\pb}{$\pi$ band}
\newcommand{\pstar}{$\pi$* band}
\newcommand{\pbs}{$\pi$ bands}
\newcommand{\kv}{\mathbf{k}}
\newcommand{\kpoint}{$\overline{\textrm{K}}$ point}
\newcommand{\kkpoint}{$\overline{\textrm{K'}}$ point}
\newcommand{\mpoint}{$\overline{\textrm{M}}$ point}
\newcommand{\kpp}{$\overline{\textrm{K'}}$}
\newcommand{\kp}{$\overline{\textrm{K}}$}
\newcommand{\mbar}{$\overline{\textrm{M}}$}
\newcommand{\gpoint}{$\overline{\textrm{\Gamma}}$ point}
\newcommand{\gk}{$\overline{\Gamma \mbox{K}}$}
\newcommand{\kpar}{$\mathbf{k}_{\parallel}$}
\newcommand{\efermi}{E$_\mathrm{F}$}
\newcommand{\edirac}{E$_\mathrm{D}$}
\newcommand{\angs}{$\mathrm{\AA}$}
\newcommand{\kk}  {$\mathbf{k}_{\overline{\textrm{K}}}$}
\newcommand{\kvec}{\underline{k}}
\newcommand{\sbs}[2]{\rlap{\textsuperscript{{#1}}}\textsubscript{#2}}
\newcommand{\round}[1]{\left({#1}\right)}
\newcommand{\moire}{moir{\'e}}
\newcommand{\gm}{$\overline{\Gamma}$}

\newcommand{\red}{\textcolor[rgb]{1,0,0}}

\newcommand{\todo}[1]{\textsl{\textcolor{red}{#1}}}

\newcommand{\kay}{\mathbf{k}}
\newcommand{\cue}{\mathbf{q}}
\newcommand{\up}{\uparrow}
\newcommand{\down}{\downarrow}
\newcommand{\str}{^{*}}
\newcommand{\pathI}{\int D[c\str,c]}

\tikzset{
        vertex/.style={rectangle,draw=black!100,fill=white!20,minimum size=8mm,thick},
        fullvertex/.style={rectangle,draw=black!100,fill=black!20,minimum size=8mm,thick},
        creation/.style={circle,fill=black!100,draw=black!100,minimum size=0mm},
        loop/.style={circle,fill=white!100,draw=black!100,minimum size=8mm,thick,->},
        annihilation/.style={circle,fill=white!20,draw=black!100,minimum size=0.5mm},
        fermion/.style={->,>=latex',thick},
        ->-/.style={decoration={  markings,  mark=at position .5 with {\arrow[scale=1,>=latex]{>}}},postaction={decorate}},
        -->-/.style={decoration={  markings,  mark=at position .75 with {\arrow[scale=1,>=latex]{>}}},postaction={decorate}},
}

% Include your paper's title here

\title{Supplementary information to 'Introducing strong correlation effects into graphene by gadolinium intercalation'}

\author{S. Link}
\affiliation{Max-Planck-Institut f\"{u}r Festk\"{o}rperforschung, Heisenbergstr. 1, D-70569 Stuttgart, Germany}

\author{S. Forti}
\altaffiliation[present address:~]{Center for Nanotechnology Innovation CNI@NEST, Piazza San Silvestro 12, 56127 Pisa, Italy}
\affiliation{Max-Planck-Institut f\"{u}r Festk\"{o}rperforschung, Heisenbergstr. 1, D-70569 Stuttgart, Germany}

\author{A. St\"{o}hr}
\affiliation{Max-Planck-Institut f\"{u}r Festk\"{o}rperforschung, Heisenbergstr. 1, D-70569 Stuttgart, Germany}

\author{K. K\"{u}ster}
\affiliation{Max-Planck-Institut f\"{u}r Festk\"{o}rperforschung,
Heisenbergstr. 1, D-70569 Stuttgart, Germany}

\author{M. R\"{o}sner}
\affiliation{Institute for Theoretical Physics, Bremen Center for
Computational Materials Science, University of Bremen Otto-Hahn-Allee 1,
28359 Bremen, Germany}
\affiliation{Department of Physics and Astronomy,
University of Southern California, Los Angeles, CA 90089-0484, USA}
\affiliation{Radboud University, Institute for Molecules \& Materials,
Heijendaalseweg 135, 6525 AJ Nijmegen, Netherlands}

\author{D. Hirschmeier}
\affiliation{Universit\"{a}t Hamburg, Institut f\"{u}r Theoretische Physik, D-20355 Hamburg, Germany}

\author{C. Chen}
\affiliation{Synchrotron SOLEIL and Universit\'e Paris Saclay, Orme des Merisiers, Saint-Aubin-BP 48, 91192 Gif sur Yvette, France}

\author{J. Avila}
\affiliation{Synchrotron SOLEIL and Universit\'e Paris Saclay, Orme des Merisiers, Saint-Aubin-BP 48, 91192 Gif sur Yvette, France}

\author{M.C. Asensio}
\affiliation{Madrid Institute of Materials Science (ICMM), Spanish Scientific
Research Council (CSIC), Cantoblanco, E-28049 Madrid - SPAIN}

\author{A.A. Zakharov}
\affiliation{MAX IV Laboratory, Lund University, Fotongatan 2, 22484 Lund,
Sweden}

\author{T.O. Wehling}
\affiliation{Institute for Theoretical Physics, Bremen Center for
Computational Materials Science, University of Bremen Otto-Hahn-Allee 1,
28359 Bremen, Germany}

\author{A.I. Lichtenstein}
\affiliation{Universit\"{a}t Hamburg, Institut f\"{u}r Theoretische Physik, D-20355 Hamburg, Germany}

\author{M.I. Katsnelson}
\affiliation{Radboud University, Institute for Molecules \& Materials, Heijendaalseweg 135, 6525 AJ Nijmegen, Netherlands}

\author{U. Starke}%
 \email[To whom correspondence should be addressed:\\ Email: ]{u.starke@fkf.mpg.de}
 \homepage[\\Electronic adress: ]{http://www.fkf.mpg.de/ga}
\affiliation{Max-Planck-Institut f\"{u}r Festk\"{o}rperforschung, Heisenbergstr. 1, D-70569 Stuttgart, Germany}

%%%%%%%%%%%%%%%%% END OF PREAMBLE %%%%%%%%%%%%%%%%

\maketitle

\setcounter{figure}{0}
\renewcommand{\thefigure}{S\arabic{figure}}

\newpage

\section{Graphene growth on SiC(0001): Buffer layer (ZLG) and quasi-free standing graphene}

Epitaxial graphene on SiC(0001) is commonly prepared by means of silicon sublimation via annealing of
the SiC substrate~\cite{Berger2004,OhtaScience2006,Riedl2007}.
An initial carbon layer develops -- well ordered in a {\sixroot} superstructure
on the SiC(0001) substrate -- which structurally is composed of the typical carbon
honeycomb lattice found in graphene. However, about one third of the
carbon atoms are still covalently bound to the topmost silicon atoms in the
substrate so that the delocalized {\pb} system cannot develop \cite{Emtsev2008,RiedlAPL2008}. This initial
carbon layer is called buffer layer or zero-layer graphene (ZLG). The
covalent bonding situation is sketched in Fig.\ \ref{S-models}(a) with the covalent bonds
between Si and C across the interface and dangling bonds (DB) on the remaining Si atoms indicated.
We note that
this bonding induces a strong buckling in the graphitic layer and leads to a
very distinct and intense quasi-(6$\times$6)-SiC(0001) diffraction
pattern~\cite{Riedl2007}.
By further annealing of the SiC sample a second carbon layer is formed, which in turn
assumes the role of the buffer layer. The initial ZLG transforms into a real
graphene layer on top with fully developed $\pi$ bands and the system is then called monolayer
graphene (MLG) (see Fig.\ \ref{S-models}(b) for a side view sketch).
{\begin{figure*}[b]
	\begin{center}
		\includegraphics[width=1.0\textwidth]{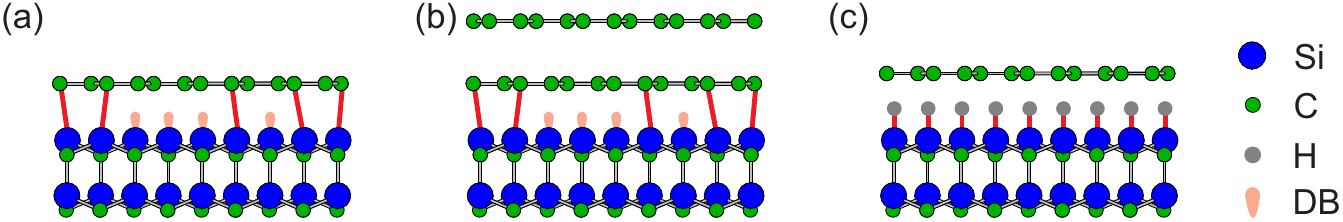}
	\end{center}
	\caption{
		Model sketches of epitaxial graphene on SiC(0001) in side view:
        (a) Buffer layer or zero-layer graphene (ZLG) with covalent Si-C interface bonds and substrate Si dangling bonds (DB),
        (b) Monolayer graphene (MLG) composed of one ZLG at the SiC interface and one van-der Waals bound graphene layer on top and
        (c) quasi-free standing graphene after hydrogen intercalation.}
	\label{S-models}
\end{figure*}}
In the present study, ZLG samples with homogeneous coverage were prepared on 6H-SiC(0001) (on-axis,
n-doped, purchased from SiCrystal GmbH) by annealing in a home-built RF-furnace in Ar
atmosphere, which results in a superb homogeneity on a waver
scale~\cite{Emtsev2009,Forti2014,Ostler2010}. Atomic force microscopy (AFM),
low energy electron diffraction (LEED) and X-ray photoelectron spectroscopy (XPS) were used to control the quality of the growth.
The ZLG carbon layer (which does not display $\pi$ bands) can be decoupled from the SiC substrate by an intercalated atomic layer, as first shown
for hydrogen intercalation~\cite{Riedl2009}. The covalent bonds at the ZLG interface are broken
and all Si atoms in the topmost substrate layer are saturated by hydrogen atoms.
Effectively in this way, a quasi-free standing graphene monolayer (QFMLG) is obtained
as depicted in the sketch in Fig.\ \ref{S-models}(c) and this QFMLG now has sharp monolayer $\pi$ bands~\cite{Riedl2009}.

\section{Intercalation method and Experimental techniques}

In the present study, decoupling by intercalation could also be achieved using a Gd layer.
In detail, Gd was evaporated onto the samples using a commercial e-beam
evaporator. During evaporation, the sample temperature is held at
800\,$^{\circ}$C. After this step, the sample is fully intercalated with Gd,
which was verified by means of XPS, LEED and angle-resolved photoelectron
spectroscopy (ARPES). With the temperature of only 800\,$^{\circ}$C during the intercalation process,
it is ensured, that no further Si sublimation takes place and no additional carbon layer is grown~\cite{Riedl2007}.
Yet, the characteristic
($13\sqrt{3}$$\times$$13\sqrt{3}$)$\textup{R}30^\circ$-graphene structure
does not yet develop. To achieve this superior order, the samples had to be heated for a very short
time (about 5~s) to about 1200\,$^{\circ}$C. The spectroscopic footprints of the
intercalated system are not affected by this treatment, i.e.~ARPES and XPS characterization is unchanged,
so that it is clear that we still deal with a quasi-free monolayer system.
After this procedure, no
degeneration could be detected when heating up to temperatures higher than
1000\,$^{\circ}$C. Also, no degeneration in the samples can be found when
keeping the samples in UHV conditions ($p<~1\times 10^{-8}$~mbar) on a
timescale of weeks. Direct exposure to air, however, leads to notable
degeneration (see below). In order to prevent degeneration during transport
to the synchrotron facilities for the ARPES measurements shown, these samples
were covered with about 100 nm of Bi which in turn was removed by heating to
temperatures higher than 700 {\cgrad} in the beamlines' vacuum facilities.
Comparative measurements by means of XPS, LEED and ARPES in the home
laboratory confirmed the validity of this treatment, and indeed no
differences to the pristine intercalated samples were found.

\section{ARPES raw data}
{\begin{figure*}[b]
	\begin{center}
		\includegraphics[width=1.0\textwidth]{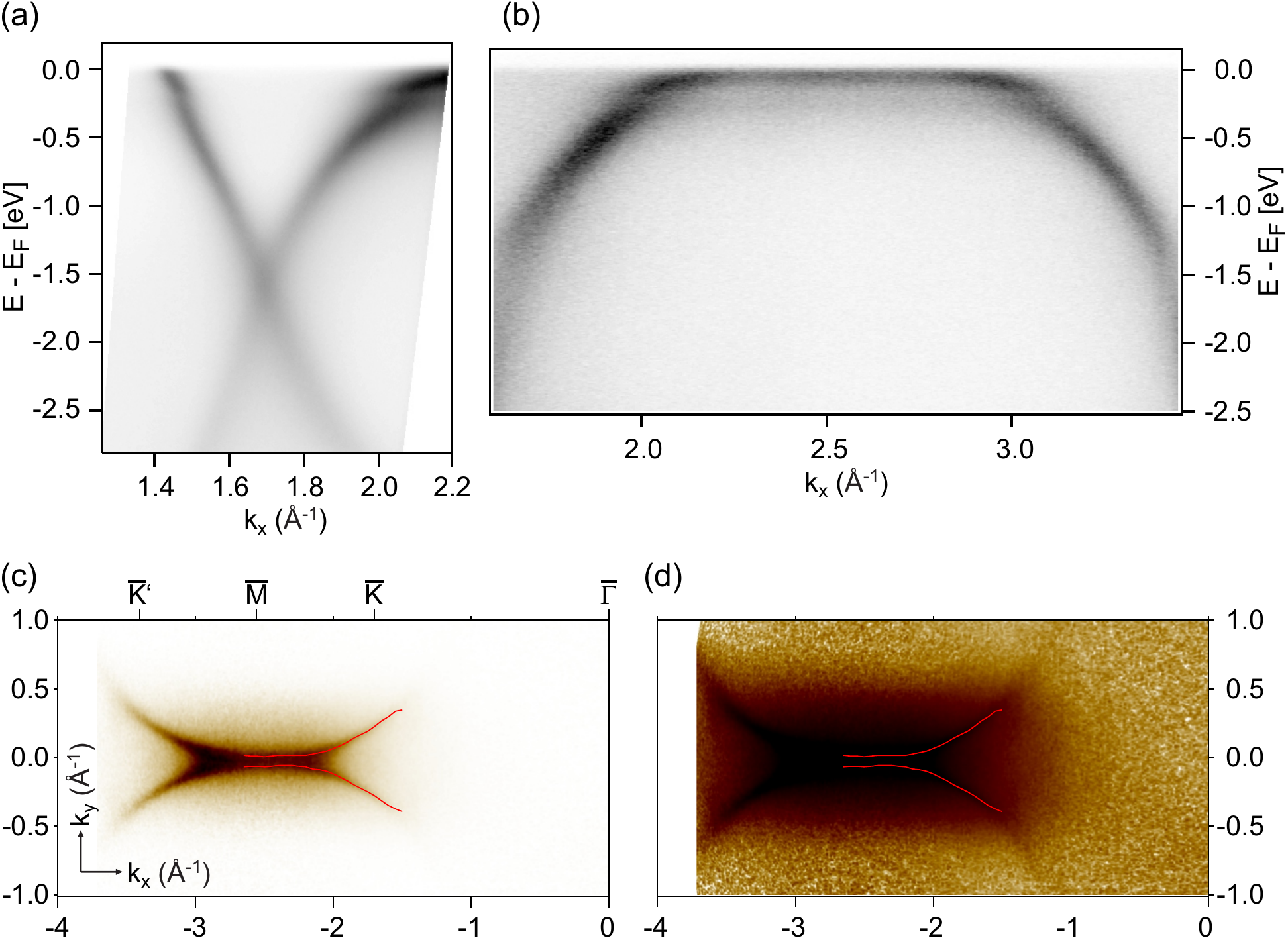}
	\end{center}
	\caption{ARPES raw data of the Gd intercalated ZLG system:
        (a) Dispersion across the {\kp} point projected along the {\gkdir} measured with 30 eV photon energy. %(data VB30{\_29})
        (b) Dispersion along the {\kp}{\mbar}{\kpp} line measured with 100 eV photon energy. %(data VB100{\_}46)
		(c) Fermi surface from the {\gm} point in the first BZ up to the {\kpp} point in the repeated BZ and
        (d) the same data set plotted with enhanced contrast and logarithmic intensity scale.
        Measurement taken with 90 eV photon energy, emission angle scan range of 55{\grad}.
        }
	\label{rawARPES}
\end{figure*}
The extended ARPES dispersion plot shown in Fig.\ 1(c) of the main text had
to be concatenated from two measurements with different photon energy for two
reasons: On the one hand, the high energy of 100 eV is necessary to allow the
full scan along the {\kp}{\mbar}{\kpp} line  since the emission angle for the
{\kpp} point would be too large at lower photon energies. On the other hand,
the low energy of 30 eV is needed for the region around the {\kp} point in
order to access both, the {\pb} and {\pstar} branches, the latter of which
would be suppressed at 100 eV photon energy due to matrix element
effects~\cite{GierzPRBR2011}. For clarity, we display the raw data used for
Fig.\ 1(c) in Fig.\ \ref{rawARPES}(a and b).
The two data sets are acquired from the same sample in the same preparation run (beamline Antares at Soleil). %sample 150424{\_}27
Figure 1(d) of the main text presents a symmetrized FS taken with 90~eV photon energy.
The FS is extracted from an exhaustive scan of ARPES dispersion slices
covering the first BZ from the {\gm} point to the {\kpoint} and continuing along the border
of the repeated BZ along the {\kp}{\mbar}{\kpp} line, thus covering a large range of emission angles (beamline I4 at MAX-Lab). %sample 141001{\_}01
In Fig.\ 1(d), the data is repeated in 60{\grad} rotation, so that the course
of the {\pbs} can be observed. We display the raw Fermi surface data
extracted from this measurement in Fig.\ \ref{rawARPES}(c) together with the
fitted band course at the Fermi energy (The Fermi surface points were
evaluated individually from momentum distribution curves extracted
perpendicular to the {\kp}{\mbar}{\kpp} line and symmetry repeated around the
first BZ in Fig.\ 1(d)). Notably, spectral weight is completely absent inside
the {\pb} ring in the first BZ, and in particular in the region of the {\gm}
point indicating that no interlayer Gd band is present in this system. This
can be seen in a contrast enhanced display of the same data set with
logarithmic intensity scale in Fig.\ \ref{rawARPES}(d). We note again, that
due to matrix element effects, the {\pstar} intensity is strongly surpressed
at the photon energy utilized in this measurement~\cite{GierzPRBR2011}. All
ARPES data were acquired with the sample liq.\ N{$_2$} cooled ({$\approx$}
90K) and using nominally linear polarization (p).

\section{Further experimental characterization}

To shine light on the state of the Gd intercalated samples we firstly
performed XPS measurements using Al K$_{\alpha}$ radiation at normal and
grazing emission for different surface sensitivity. Fig.\ \ref{S1}(a) shows
the spectra related to the C~1s, Si~2p and Gd~3d$_{5/2}$ core levels
respectively. The relative intensity is normalized to the Gd~3d peak at
0{\grad} and 60{\grad} respectively.
\begin{figure*}
	\begin{center}
		\includegraphics[width=1.0\textwidth]{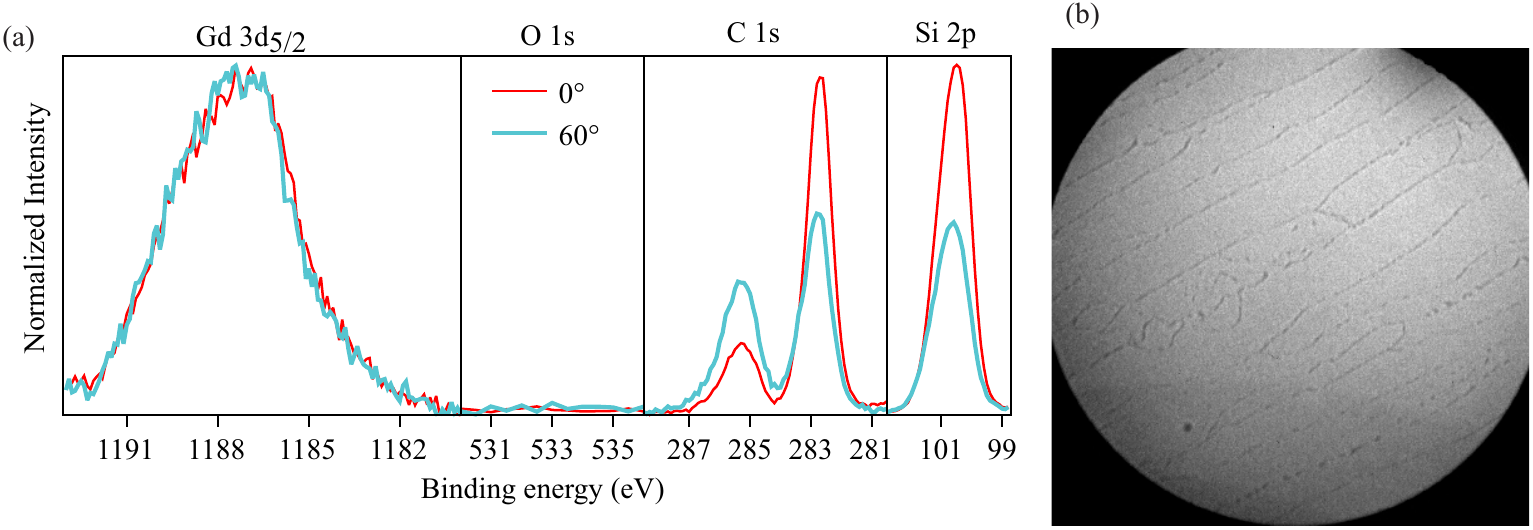}
	\end{center}
	\caption{
		(a) XPS measurements in the binding energy regions of the Gd~3d, O~1s, C~1s,
		Si~2p acquired with Al K$_{\alpha}$ radiation at different emission angles. (b) LEEM
		micrograph of a Gd intercalated ZLG sample with a field of view of 25~$\mu$m measured with electron energy 1.2~eV.}
	\label{S1}
\end{figure*}
It is well visible, that the Si~2p component as well as the C~1s component at
$282.7 \textup{eV}$ binding energy are suppressed when increasing the emission
angle. The relative intensity of the C1s component at about $285 \textup{eV}$
binding energy, however, increases when increasing the emission angle. The
LEEM micrograph shown in Fig.\ \ref{S1}(b) displays a completely homogeneous
surface without any Gd clusters on top. The only contrast visible is related
to step edges in the SiC substrate, which is intrinsic to epitaxial graphene
on SiC(0001). Considering the simplicity of the initial ZLG system, i.e. one
graphitic layer on top of the SiC, this information is sufficient to conclude
that the Gd is sandwiched by the SiC and the graphitic layer. It has to be
noted, that no O~1s component appears for these samples when safekeeping
under UHV conditions. This is also indicative for the sandwich configuration,
as the graphene layer protects the Gd from oxidation by residual gas in the
vacuum chamber.
\begin{figure*}
	\begin{center}
		\includegraphics[width=1.0\textwidth]{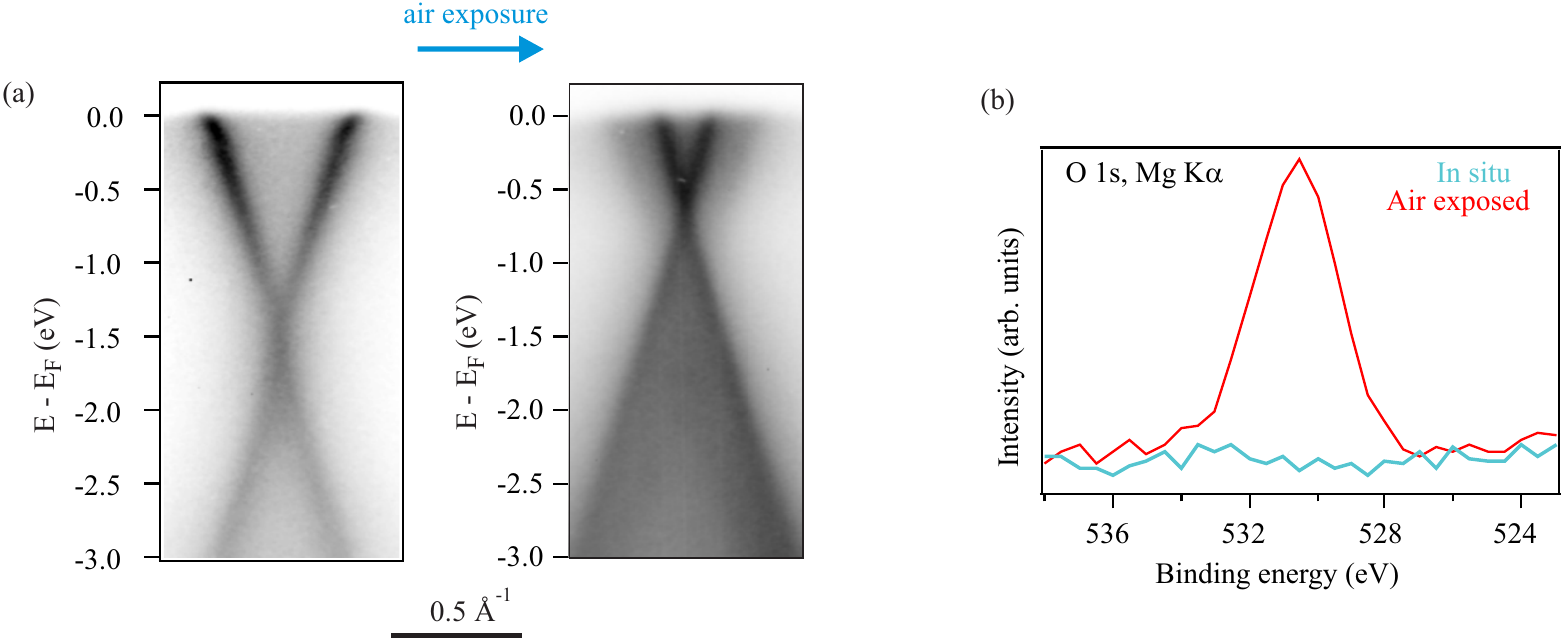}
	\end{center}
	\caption{(a) ARPES on the $\pi$ bands before and after air exposure. (b) XPS in the region of the O~1s level before and after air exposure.}
	\label{S2}
\end{figure*}
Direct exposure to air induces degradation. As visible in the ARPES cuts in
Fig.\ \ref{S2}(a) (photon energy 60 eV, beamline I4 at MAX-Lab), the doping
level gets strongly reduced, so that the Dirac point is located at about
0.5~eV, whereas it is about 1.6~eV in the pristine intercalated samples. In
XPS a strong O~1s can then be found. This cannot be reversed by mild
outgasing in UHV to remove adsorbed water or oxygen. One can therefore
conclude that under air exposure, oxygen or water migrates beneath the
graphene layer in these samples and presumably reacts chemically with the Gd.
The XPS measurements in Fig.\ \ref{S2}(b) as well as Fig.\ 3(c,e) were
taken at beamline I311 at MAX-Lab.

\section{Density functional theory}

The density functional theory (DFT) calculations were performed within the
generalized gradient approximation (GGA) \cite{Perdew1996} using the Vienna
Ab Initio Simulation Package (VASP) \cite{Kresse1994} and the projector
augmented wave (PAW) \cite{Bloechel1994, Kresse1999} basis sets. We account
for the strong local Coulomb interactions of the Gd 4f-electrons within the
GGA+U approach using the Coulomb parameters $U=7 \,$eV and $J=1\,$eV, which
have been shown to be well suited to describe rare earth systems
\cite{Anisimov1997, Larson2007}. We considered Gd adsorbed on graphene in the
{\rootthree} structure as shown in Fig.\ 3 (c) of the manuscript. While the
graphene lattice constant was fixed to a value of $a_0 \approx 2.48\,$\AA,
the Gd $z$ position was relaxed until forces were below
$0.001\,$eV$\,$\AA$^{-1}$ resulting in a $z$ distance of $a_z \approx
2.23\,$\AA.  After finding the self-consistent ground state, we use the
BandUP code \cite{Medeiros2014, Medeiros2015} to unfold the spin-resolved
band structure of the {\rootthree} unit cell to the primitive cell of
graphene.

\section{Fluctuating exchange approximation}
The spectral function shown in panel (b) of Fig.\ 4 was calculated using the
fluctuating exchange approximation (FLEX) \cite{Bickers1989}, which will be
outlined in this section. Starting with the partition function in the path
integral representation
\begin{equation}\label{eq:partition}
\mathcal{Z}=\pathI e^{-S[c\str,c]},
\end{equation}
where the action depends on the complex Grassmann fields $c\str$ and $c$, representing the respective creation and annihilation of fermions, we investigate electronic correlation effects in doped graphene by considering the Hubbard model on the bipartite honeycomb lattice which can be written by
\begin{align}\label{eq:Hubbard_action}
	S=S_0+V=&-\sum_{ij\sigma\nu}c\str_{i\sigma\nu}\left[(i\nu_n+\mu)\delta_{ij}-\epsilon^{ij}_{\kay}\right]c_{j\sigma\nu} \notag \\
	&+U\sum_{j\nu\nu'\Omega}c^{*\up}_{j\nu}c^{\up}_{j\nu+\Omega}c^{*\down}_{j\nu'+\Omega}c^{\down}_{j\nu'},
\end{align}
where $\sigma=\up,\down$ is the spin projection, $\nu_n=(2n+1)\pi T$ the n-th fermionic Matsubara frequency, $\Omega=2n\pi T$, the n-th bosonic Matsubara frequency, $\kay$ the lattice momentum and the sum is taken over the four momentum $\nu=(i\nu_n,\kay)$. Latin letters $i,j$ represent the two sublattices of the honeycomb lattice. Note that indices being super or subscript does not hold any further meaning here and in the following and we will only use one or the other for the sake of readability. The model parameters are the chemical potential $\mu$, temperature $T$, Hubbard interaction $U$ and nearest-neighbour hopping amplitude $t$. The latter enters the tight-binding dispersion $\epsilon_{\kay}$, which can be written as a $2\times2$ matrix in sublattice degrees of freedom
\begin{equation}
\epsilon_{\kay}=
\begin{pmatrix}
0 & -t f_{\kay} \\
-t f^{*}_{\kay}& 0
\end{pmatrix},
\end{equation}
where $f_{\kay}=e^{-i( \frac{a}{\sqrt{3}} k_y)}+2e^{i (\frac{a}{2\sqrt{3}} k_y)} \cos\left( \frac{a}{2} k_x\right)$ with $a$ being the lattice constant. We set the main diagonal to zero, as we restrict ourselves to nearest-neighbour hopping. Note that within our model the interaction is restricted to a local density-density interaction and we ignore the Coulomb repulsion between electrons residing on different sites.

Theoretically the ARPES data correspond to the energy and momentum resolved spectral function given by the imaginary part of the retarded Greens function $A_{\kay,\omega}=\frac{-1}{\pi} Im G_{\kay,\omega+i0^{+}}$, where $\omega$ labels real frequencies. Correlation effects between electrons enter the Greens function through the self-energy $\Sigma_{\nu}$ via the Dyson equation
\begin{equation}\label{eq:dyson}
G_{\nu}=
\begin{pmatrix}
\nu +\mu -\Sigma^{00}_{\nu} & -\epsilon_{\kay}-\Sigma^{01}_{\nu} \\
-\epsilon^{*}_{\kay} -\Sigma^{10}_{\nu} & \nu +\mu-\Sigma^{11}_{\nu}
\end{pmatrix}^{-1}.
\end{equation}
Therefore our problem boils down to finding an appropriate approximation to the self-energy.
%As the ARPES data of the experiment shows a hint of a shadow band slightly below the flat band along the Fermi energy,
We expect that correlation effects from antiferromagnetic fluctuations are responsible for the band flatting, which can be approximated by ladder diagrams of particle-hole type. The corresponding self-energy can be written in terms of these diagrams using the Schwinger-Dyson equation (SDE), which can be derived from the equation of motion for the Greens function and the Dyson equation. In the language of the coherent state path integral formalism the SDE can be obtained from
\begin{equation}
\Sigma_{\alpha\beta}G_{\beta\gamma}=\left\langle c\str_\gamma \frac{\partial V}{\partial c\str_\alpha}\right\rangle,
\end{equation}
where greek letters represent general single-particle degrees of freedom. Inserting the interaction $V$ from the second line of Eq.~(\ref{eq:Hubbard_action}), the SDE for the problem at hand is

\begin{equation}\label{eq:hubbard_sde}
\Sigma^{\sigma ij}_{\nu}=U\langle n_{\bar{\sigma}}\rangle\delta_{ij}-U\sum_{\nu'\omega,acd}G^{\sigma ia}_{\nu+\omega}\Gamma^{\bar{\sigma}\bar{\sigma}\sigma\sigma,ajcd}G^{\bar{\sigma}di}_{\nu'+\omega}G^{\bar{\sigma}ic}_{\nu'},
\end{equation}
where $\Gamma$ is the full two-particle vertex representing all connected diagrammatic contributions to the two-particle Greens function and $\bar{\sigma}=-\sigma$. Note that we are only considering the paramagnetic case, where all quantities are invariant under rotation of spins.
The two-particle Greens function is defined as
\begin{equation}
G^{(2) abcd}_{\nu\nu'\Omega}=\frac{1}{\mathcal{Z}}\pathI c^a_{\nu+\omega} c^{* b}_{\nu} c^{c}_{\nu'} c^{* d}_{\nu'+\omega} e^{-S[c\str,c]}
\end{equation}
and is related to $\Gamma$ via
\begin{align}
	G^{(2) abcd}_{\nu\nu'\Omega}&=G^{ab}_{\nu+\Omega}G^{cd}_{\nu'+\Omega}\delta_{\Omega 0}-G^{ad}_{\nu+\Omega}G^{cb}_{\nu'+\Omega}\delta_{\nu\nu'} \notag \\
	&+G^{aa'}_{\nu+\Omega}G^{cc'}_{\nu}\Gamma^{a'b'c'd'}_{\nu\nu'\Omega}G^{b'b}_{\nu'}G^{d'd}_{\nu'+\Omega},
\end{align}
where the last equation defines the two-particle vertex. Diagrammatic contributions to the two-particle vertex fall into one of two groups, those which are fully irreducible and those which are reducible with respect to one of three channels. Those channels are the particle-particle (pp), the vertical (v) and the horizontal (h) particle-hole channel. Reducibility with respect to one of these channels means, that a diagram can be split into two diagrams by respectively cutting a particle-particle, vertical particle-hole or horizontal particle-hole pair propagator. The full two-particle vertex can thus be written as
\begin{equation}\label{eq:firr_red}
\Gamma=\Gamma_{firr}+\Gamma^{(p)}_{red}+\Gamma^{(v)}_{red}+\Gamma^{(h)}_{red},
\end{equation}
where $\Gamma_{firr}$ represents all fully irreducible contributions and $\Gamma^{(\alpha)}_{red}$ represents the contributions reducible with respect to channel $\alpha=p,v,h$. As neither the self-energy nor the two-particle vertex are known exactly for the problem defined by Eqs.~(\ref{eq:partition})-(\ref{eq:Hubbard_action}), sensible approximations, that is, suitable to qualitatively describe the physical phenomenon of interest, have to be employed. In this work an approximation to the self-energy is obtained from the SDE and a suitable approximation to the vertex which is threefold:
\begin{itemize}
	\item The fully irreducible two-particle vertex is approximated by the bare interaction $\Gamma_{firr}\approx U$.
	\item Only reducible two-particle diagrams in either the vertical or the horizontal particle-hole channel are taken into account, whereas particle-particle reducible contributions are neglected $\Gamma^{(p)}_{red}=0$.
	\item All reducible contributions are obtained from the solution of the Bethe-Salpeter equation in the respective channel $\Gamma^{(\alpha)}$, where the irreducible diagrams with respect to either channel are approximated again by the bare interaction $U$ and the corresponding reducible contribution is then $\Gamma^{(\alpha)}_{red}=\Gamma^{(\alpha)}-U$.
\end{itemize}
Thus Eq.~(\ref{eq:firr_red}) becomes $\Gamma=\Gamma^{(h)}+\Gamma^{(v)}-U$, where the indices $v$, $h$ represent the vertical and horizontal particle-hole channel respectively. The $\Gamma^{(\alpha)}$ contributions to the SDE are now given by ladder diagrams of the form
\begin{equation}\label{eq:gamma_h_uudd}
\tikz[baseline={([yshift=-6mm]current bounding box.center)}, inner sep=1,auto]{
	\node (gamma) at (0,-1) {$\mathbf{\Gamma^{(h)iijj}_{\up\up\down\down}}$};%%%%%%%%%%%%%compared to the file used in the masterthesis i have switched the lines%%%%%%%%%%%%%%%%%%%%%%%%%%%%%%%%%%
	\node (equal) at (0.8,-1) {$\mathbf{=}$};
	\node(U) at (1.2,-1.) {};
	\node(Ua) at ($(U) + (135:6mm)$) {};
	\node(Ub) at ($(U) + (225:6mm)$) {};
	\node(Uc) at ($(U) + (8mm,0)$) {};
	\node(Ud) at ($(Uc)+(45:6mm)$) {};
	\node(Ue) at ($(Uc)+(315:6mm)$) {};
	\draw (Ua) to node {$\up$i} (U) [->-];
	\draw (U) to node {$\up$i} (Ub) [->-];
	\draw[thick,dashed] (U) to (Uc);
	\draw (Ue) to node {$\down$i} (Uc) [->-];
	\draw (Uc) to node {$\down$i} (Ud) [->-];
	\node (plus) at (2.7,-1) {$\mathbf{+}$};
	\node(U1) at (3.3,-1) {};
	\node(U1a) at ($(U1) + (135:6mm)$) {};
	\node(U1b) at ($(U1) + (225:6mm)$) {};
	\node(U1c) at ($(U1) + (0:6mm)$) {};
	\draw (U1a) to node {$\up$i} (U1) [->-];
	\draw (U1) to node {$\up$i} (U1b) [->-];
	\draw[thick,dashed] (U1) to (U1c);
	\node(U2a) at ($(U1c)+(8mm,0)$) {};
	\node(U2b) at ($(U2a)+(6mm,0)$) {};
	\draw (U1c) [bend right = 60,swap] to node {$\down$} (U2a) [->-];
	\draw (U2a) [bend right = 60,swap] to node {$\down$} (U1c) [->-];
	\draw[thick,dashed] (U2a) to (U2b);
	\node(U3a) at ($(U2b)+(8mm,0)$) {};
	\node(U3b) at ($(U3a)+(6mm,0)$) {};
	\draw (U2b) [bend right = 60,swap] to node {$\up$} (U3a) [->-];
	\draw (U3a) [bend right = 60,swap] to node {$\up$} (U2b) [->-];
	\node(U3c) at ($(U3b) + (45:6mm)$) {};
	\node(U3d) at ($(U3b) + (315:6mm)$) {};
	\draw[thick,dashed] (U3a) to (U3b);
	\draw (U3b) to node {$\down$j} (U3c) [->-];
	\draw (U3d) to node {$\down$j} (U3b) [->-];
	\node (pluss) at (7.4,-1) {$\mathbf{+}$};
	\node (dotdot) at (8.,-1) {...};
}
\end{equation}

\begin{equation}\label{eq:gamma_v_uudd}
%%%%%%%%%%%%%%%%%%%%%%%%%%%%%%%%%%%%%%%%%%%%%%%%%%%%%%%%%%%%%%%%%%%%%%
%%%%%%%%%%%%%%%%%%%%%%%%Gammavupupdowndown%%%%%%%%%%%%%%%%%%%%%%%%%%%%
\tikz[baseline={([yshift=-6mm]current bounding box.center)}, inner sep=1,auto]{
	\node (gamma) at (0,-1) {$\mathbf{\Gamma^{(v)iijj}_{\up\up\down\down}}$};%%%
	\node (equal) at (0.8,-1) {$\mathbf{=}$};
	\node(U) at (1.2,-1.) {};
	\node(Ua) at ($(U) + (135:6mm)$) {};
	\node(Ub) at ($(U) + (225:6mm)$) {};
	\node(Uc) at ($(U) + (8mm,0)$) {};
	\draw[thick,dashed] (U) to (Uc);
	\node(Ud) at ($(Uc)+(45:6mm)$) {};
	\node(Ue) at ($(Uc)+(315:6mm)$) {};
	\draw (Ua) to node {$\up$i} (U) [->-];
	\draw (U) to node {$\up$i} (Ub) [->-];
	\draw (Ue) to node {$\down$i} (Uc) [->-];
	\draw (Uc) to node {$\down$i} (Ud) [->-];
	\node (plus) at (2.7,-1) {$\mathbf{+}$};
	%%%%%%%%%%%%%%%%%%second order%%%%%%%%%%%%%%%%%%%%%%%%%%%%%%%%%%%%%%%%
	\node(U1) at (3.6,0) {};
	\node(U1a) at ($(U1) + (-4mm,0mm)$) {};
	\node(U1b) at ($(U1) + (+4mm,0mm)$) {};
	\node(U1c) at ($(U1a) + (0mm,-6mm)$) {};
	\node(U1d) at ($(U1b) + (0mm,-6mm)$) {};
	\draw[thick,dashed] (U1c) to (U1d);
	\node(U2a) at ($(U1c) + (0mm,-6mm)$) {};
	\node(U2b) at ($(U1d) + (0mm,-6mm)$) {};
	\draw[thick,dashed] (U2a) to (U2b);
	\node(U2c) at ($(U2a) + (0mm,-6mm)$) {};
	\node(U2d) at ($(U2b) + (0mm,-6mm)$) {};
	\draw(U1a) [swap] to node {$\up$i} (U1c) [->-];
	\draw(U1c) [swap] to node {$\up$} (U2a) [->-];
	\draw(U2a) [swap] to node {$\up$j} (U2c) [->-];
	\draw(U2d) [swap] to node {$\down$j} (U2b) [->-];
	\draw(U2b) [swap] to node {$\down$} (U1d) [->-];
	\draw(U1d) [swap] to node {$\down$i} (U1b) [->-];
	%%%%%%%%%%%%%%%%%%%%%%%%%%%%%%%%%%%%%%%%%%%%%%%%%%%%%%%%%%%%%%%%%%%%%%
	\node (plus) at (4.5,-1) {$\mathbf{+}$};
	%%%%%%%%%%%%%%%%%%%third order%%%%%%%%%%%%%%%%%%%%%%%%%%%%%%%%%%%%%%%%
	\node(U3) at (5.5,0.2) {};
	\node(U3a) at ($(U3) + (-4mm,0mm)$) {};
	\node(U3b) at ($(U3) + (+4mm,0mm)$) {};
	\node(U3c) at ($(U3a) + (0mm,-6mm)$) {};
	\node(U3d) at ($(U3b) + (0mm,-6mm)$) {};
	\draw[thick,dashed] (U3c) to (U3d);
	\node(U4a) at ($(U3c) + (0mm,-6mm)$) {};
	\node(U4b) at ($(U3d) + (0mm,-6mm)$) {};
	\draw[thick,dashed] (U4a) to (U4b);
	\node(U4c) at ($(U4a) + (0mm,-6mm)$) {};
	\node(U4d) at ($(U4b) + (0mm,-6mm)$) {};
	\draw[thick,dashed] (U4c) to (U4d);
	\node(U5a) at ($(U4c) + (0mm,-6mm)$) {};
	\node(U5b) at ($(U4d) + (0mm,-6mm)$) {};
	\node(U5c) at ($(U5a) + (0mm,-6mm)$) {};
	\node(U5d) at ($(U5b) + (0mm,-6mm)$) {};
	\draw(U3a) [swap] to node {$\up$i} (U3c) [->-];
	\draw(U3c) [swap] to node {$\up$} (U4a) [->-];
	\draw(U4a) [swap] to node {$\up$} (U4c) [->-];
	\draw(U4c) [swap] to node {$\up$j} (U5a) [->-];
	\draw(U5b) [swap] to node {$\down$j} (U4d) [->-];
	\draw(U4d) [swap] to node {$\down$} (U4b) [->-];
	\draw(U4b) [swap] to node {$\down$} (U3d) [->-];
	\draw(U3d) [swap] to node {$\down$i} (U3b) [->-];
	\node (pluss) at (6.5,-1) {$\mathbf{+}$};
	\node (dotdot) at (7.,-1) {... .};
}
\end{equation}

Note that the restriction to ladder diagrams only generates reducible vertex contributions with sublattice structure $\Gamma^{ijkl}=\Gamma^{iijj}\delta_{ij}\delta_{kl}$. While the vertex components contributing to the SDE come from both the vertical and horizontal channel and have spin structure $\up\up\down\down$, it is possible to evaluate the SDE using only the horizontal channel by exploiting symmetries. First note that, up to a sign, Eq.~(\ref{eq:gamma_v_uudd}) is equivalent to
\begin{equation}\label{eq:gamma_h_uddu}
\tikz[baseline={([yshift=-6mm]current bounding box.center)}, inner sep=1,auto]{
	%%%%%%%%%%%%%%%%%%%%%%%%%%%%%%%%%%%%%%%%%%%%%%%%%%%%%%%%%%%%%%%%%%%%%%%%%%%%%%%%%%%%%
	\node (gamma) at (0,-1) {$\mathbf{\Gamma^{(h)ijji}_{\up\down\down\up}}$};%%%%%%%%
	\node (equal) at (0.8,-1) {$\mathbf{=}$};
	%%%%%%%%%%%%%%%%%%%%%%%%%first order%%%%%%%%%%%%%%%%%%%%%%%%%%%%%%%%%%%%%%
	\node (gamma1) at(2.1,-1)  {};
	\node (a1) at ($(gamma1) + (-4mm,4mm)$) {};
	\node (a2) at ($(gamma1) + (-4mm,-4mm)$) {};
	\node (a3) at ($(a1)+(-6mm,0mm)$) {};
	\node (a4) at ($(a1)+(6mm,0mm)$) {};
	\node (a5) at ($(a2)+(-6mm,0mm)$) {};
	\node (a6) at ($(a2)+(6mm,0mm)$) {};
	\draw[thick,dashed] (a1) to (a2);
	\draw (a3) to node {$\up$i} (a1) [->-];
	\draw (a1) to node {$\up$i} (a4) [->-];
	\draw (a2) to node {$\down$i} (a5) [->-];
	\draw (a6) to node {$\down$i} (a2) [->-];
	%%%%%%%%%%%%%%%%%%%%%%%%%%%%%%%%%%%%%%%%%%%%%%%%%%%%%%%%%%%%%%%%%%%%%%%%%%
	\node (plus) at (2.4,-1) {$\mathbf{+}$};
	%%%%%%%%%%%%%%%second order%%%%%%%%%%%%%%%%%%%%%%%%%%%%%%%%%%%%%%%%%%%%%%%
	\node (gamma2) at(3.5,-1) {};
	\node (b1) at ($(gamma2) + (-4mm,4mm)$) {};
	\node (b2) at ($(gamma2) + (-4mm,-4mm)$) {};
	\node (b3) at ($(b1)+(-6mm,0mm)$) {};
	\node (b4) at ($(b2)+(-6mm,0mm)$) {};
	\draw[thick,dashed] (b1) to (b2);
	\node (gamma3) at(4.,-1) {};
	\node (c1) at ($(gamma3) + (-4mm,4mm)$) {};
	\node (c2) at ($(gamma3) + (-4mm,-4mm)$) {};
	\draw[thick,dashed] (c1) to (c2);
	\node (c3) at ($(c1)+(6mm,0mm)$) {};
	\node (c4) at ($(c2)+(6mm,0mm)$) {};
	\draw (b3) to node {$\up$i} (b1) [->-];
	\draw (b1) to node {$\up$} (c1) [->-];
	\draw (c1) to node {$\up$j} (c3) [->-];
	\draw (c4) to node {$\down$j} (c2) [->-];
	\draw (c2) to node {$\down$} (b2) [->-];
	\draw (b2) to node {$\down$i} (b4) [->-];
	%%%%%%%%%%%%%%%%%%%%%%%%%%%%%%%%%%%%%%%%%%%%%%%%%%%%%%%%%%%%%%%%%%%%%%%%%%
	\node (pluss) at (4.6,-1) {$\mathbf{+}$};
	%%%%%%%%%%%%%%third order%%%%%%%%%%%%%%%%%%%%%%%%%%%%%%%%%%%%%%%%%%%%%%%%%
	\node (gamma5) at(5.7,-1) {};
	\node (e1) at ($(gamma5) + (-4mm,4mm)$) {};
	\node (e2) at ($(gamma5) + (-4mm,-4mm)$) {};
	\node (e3) at ($(e1)+(-6mm,0mm)$) {};
	\node (e4) at ($(e2)+(-6mm,0mm)$) {};
	\draw[thick,dashed] (e1) to (e2);
	\node (gamma6) at(6.2,-1) {};
	\node (f1) at ($(gamma6) + (-4mm,4mm)$) {};
	\node (f2) at ($(gamma6) + (-4mm,-4mm)$) {};
	\draw[thick,dashed] (f1) to (f2);
	\node (gamma7) at(6.7,-1) {};
	\node (g1) at ($(gamma7) + (-4mm,4mm)$) {};
	\node (g2) at ($(gamma7) + (-4mm,-4mm)$) {};
	\draw[thick,dashed] (g1) to (g2);
	\node (g3) at ($(g1)+(6mm,0mm)$) {};
	\node (g4) at ($(g2)+(6mm,0mm)$) {};
	\draw (e3) to node {$\up$i} (e1) [->-];
	\draw (e1) to node {$\up$} (f1) [->-];
	\draw (f1) to node {$\up$} (g1) [->-];
	\draw (g1) to node {$\up$j} (g3) [->-];
	\draw (g4) to node {$\down$j} (g2) [->-];
	\draw (g2) to node {$\down$} (f2) [->-];
	\draw (f2) to node {$\down$} (e2) [->-];
	\draw (e2) to node {$\down$i} (e4) [->-];
	%%%%%%%%%%%%%%%%%%%%%%%%%%%%%%%%%%%%%%%%%%%%%%%%%%%%%%%%%%%%%%%%%%%%%%%%%%
	\node (pluss) at (7,-1) {$\mathbf{+}$};
	\node (dotdot) at (7.5,-1) {$...$};
}
\end{equation}

where even as well as odd orders contribute. While the odd orders are, again up to a sign, equivalent to the contributions from Eq.~(\ref{eq:gamma_h_uudd}) the even orders are equivalent to the diagrams given by
\begin{equation}\label{eq:gamma_h_uuuu}
\tikz[baseline={([yshift=-6mm]current bounding box.center)}, inner sep=1,auto]{
	\node (gamma) at (0,-1) {$\mathbf{\Gamma^{(h)iijj}_{\up\up\up\up}}$};%%%%%%%%%%%%%compared to the file used in the masterthesis i have switched the lines%%%%%%%%%%%%%%%%%%%%%%%%%%%%%%%%%%
	\node (equal) at (0.8,-1) {$\mathbf{=}$};
	\node(U1) at (1.2,-1) {};
	\node(U1a) at ($(U1) + (135:6mm)$) {};
	\node(U1b) at ($(U1) + (225:6mm)$) {};
	\node(U1c) at ($(U1) + (0:6mm)$) {};
	\draw (U1a) to node {$\up$i} (U1) [->-];
	\draw (U1) to node {$\up$i} (U1b) [->-];
	\draw[thick,dashed] (U1) to (U1c);
	\node(U2a) at ($(U1c)+(8mm,0)$) {};
	\node(U2b) at ($(U2a)+(6mm,0)$) {};
	\draw (U1c) [bend right = 60,swap] to node {$\down$} (U2a) [->-];
	\draw (U2a) [bend right = 60,swap] to node {$\down$} (U1c) [->-];
	\draw[thick,dashed] (U2a) to (U2b);
	\node(U3a) at ($(U2b)+(45:6mm)$) {};
	\node(U3b) at ($(U2b)+(315:6mm)$) {};
	\draw (U2b) to node {$\up$j} (U3a) [->-];
	\draw (U3b) to node {$\up$j} (U2b) [->-];
	\node(U3c) at ($(U3b) + (45:6mm)$) {};
	\node(U3d) at ($(U3b) + (315:6mm)$) {};
	\node (plus) at (4.,-1) {$\mathbf{+}$};
	\node (dotdot) at (4.7,-1) {$...\ .$};
}
\end{equation}

Hence $\Gamma^{\up\up\down\down}=2\Gamma^{(h)\up\up\down\down}-\Gamma^{(h)\up\up\up\up}-U$. The dots at the end of Eqs.~(\ref{eq:gamma_h_uudd})-(\ref{eq:gamma_h_uuuu}) indicate the summation of an infinite power series in $U$ and yield
\begin{align}
	\Gamma^{(h)\up\up\down\down}&=U\sum_{n}\left(U^{2}\chi_{\omega}^{2}\right)^{n}=U\left(1-U^{2}\chi_{\omega}^{2}\right)^{-1} \\
	\Gamma^{(h)\up\up\up\up}&=-U^{2}\chi_{\omega}\sum_{n}\left(U^{2}\chi_{\omega}^{2}\right)^{n}=-U^{2}\chi\left(1-U^{2}\chi_{\omega}^{2}\right)^{-1} \\
	\Gamma^{(h)\up\down\down\up}&=-U\sum_{n}\left(U\chi_{\omega}\right)^{n}=-U\left(1-U\chi_{\omega}\right)^{-1}\\
	\chi_{\omega}&=\frac{1}{N}\sum_{\nu}G_{\nu+\omega}G_{\nu},
\end{align}
and it is easy to check that $\Gamma^{(h)\up\down\down\up}=\Gamma^{(h)\up\up\up\up}-\Gamma^{(h)\up\up\down\down}$. Note that the dependence on sublattice degrees of freedom was ignored in the equations above. However it can be easily restored by assigning matrix value to the left and right hand side of the equations and respectively replace the interaction $U$ and the particle-hole bubble with $2\times2$ matrices in sublattice space according to
\begin{align}
	\hat{U}&=U\mathbf{1} \\
	\hat{\chi}&=\begin{pmatrix}
		\chi^{00} & \chi^{01} \\
		\chi^{10} & \chi^{11}
	\end{pmatrix} \\
	\chi^{ab}_{\omega}&=\frac{1}{N}\sum_{\nu}G^{ab}_{\nu+\omega}G^{ba}_{\nu}
\end{align}
Thus dropping the spin dependence one ends up with the final expression for the self-energy
\begin{equation}\label{eq:sde_final}
\Sigma^{ij}_{\nu}=U\langle n_{\bar{\sigma}}\rangle\delta_{ij}-U\sum_{\omega,k}G^{ij}_{\nu+\omega}\Gamma^{jjkk}\chi^{ki}_\omega,
\end{equation}
where $\Gamma^{jjkk}$ is given by the $j,k$ component of matrix
\begin{equation}
\hat{\Gamma}=\frac{3}{2}U\left(1+U\hat{\chi}\right)^{-1}+\frac{U}{2}\left(1-U\hat{\chi}\right)^{-1}
\end{equation}
Using the thus obtained self-energy in the Dyson equation (\ref{eq:dyson}) we
gain the renormalized Greens function $G_{\nu}$ from which we obtain the
spectral function shown in Fig.\ 4 (b) of the main text.

All equations are evaluated using the bare propagators.